\documentclass[10pt]{article}
\usepackage{epsf}
\renewcommand{\baselinestretch}{2}
\usepackage{anysize}
\usepackage{graphicx}
\usepackage{amssymb,amsmath}
\marginsize{1 in}{1 in}{1 in}{1 in}
\newcommand{\ms}{\medspace}
\newcommand{\gr}{$\mathcal{G}$}

\newcommand{\tr}{{\rm Tr}}
\newcommand{\e}{\varepsilon}
\newcommand{\sd}{Schr\"{o}dinger }
\newcommand{\curv}{$\mathcal{C}$}
\newcommand{\U}{\mathcal{U}}
\newcommand{\norm}[1]{\left|\left|#1\right|\right|}

\begin{document}

\title{Search complexity and resource scaling for the quantum optimal control of unitary transformations}
\author{Katharine W. Moore, Raj Chakrabarti\footnote{Current address: Department of Chemical Engineering, Purdue University, West Lafayette, IN, 47907}, Gregory Riviello, and Herschel Rabitz\\
Department of Chemistry, Princeton University, Princeton, NJ, 08544}
\maketitle
\begin{abstract}
The optimal control of unitary transformations is a fundamental problem in quantum control theory and quantum information processing. The feasibility of performing such optimizations is determined by the computational and control resources required, particularly for systems with large Hilbert spaces. Prior work on unitary transformation control indicates that (i) for controllable systems, local extrema in the search landscape for optimal control of quantum gates have null measure, facilitating the convergence of local search algorithms; but (ii) the required time for convergence to optimal controls can scale exponentially with Hilbert space dimension. Depending on the control system Hamiltonian, the landscape structure and scaling may vary. This work introduces methods for quantifying Hamiltonian-dependent and kinematic effects on control optimization dynamics in order to classify quantum systems according to the search effort and control resources required to implement arbitrary unitary transformations. 
\end{abstract}

\section{Introduction}

The methodology of optimal control theory (OCT) has been applied to achieve various dynamical objectives in quantum systems by manipulating constructive quantum wave interference to maximize the likelihood of attaining desired target states \cite{Rice2000}. Three classes of problems - state control, observable control, and unitary transformation or gate control \cite{constantin} -  have received the most attention in the quantum control community to date.
The generation of targeted unitary transformations is of both fundamental interest and has direct applications to quantum information sciences since the quantum logic gates required to carry out quantum computation are represented by unitary transformations \cite{NielsenChuang}.
Since $U(N)$ (and $SU(N)$) are compact Lie groups, it is possible to generate  any $U \in U(N)$ through sequential application of elements of a complete set of generators 
$\imath H_1, \cdots, \imath H_k$ for $U(N)$, i.e., $W = \exp(\imath H_k t_k) \cdots \exp(\imath H_1  t_1)$. This strategy (\emph{uniform finite generation}) is now commonly applied in gate decomposition strategies wherein the unitary gate $W$ expressed in terms of $n$ qubits (i.e., with a corresponding $2^n$ dimensional Hilbert space) is constructed through sequential application of various $U_j = \exp( \imath H_j t_j)$ which each act on only 1-2 qubits \cite{DAlessandro,NielsenChuang}. However, provided the system is controllable, it is also possible to generate any $W$ by shaping time-dependent \emph{control functions} $\e_j(t),~j=2,\cdots,k$ over a time interval $[0,T]$.
Each control function is coupled to a corresponding control Hamiltonian $\imath H_j,~j=2,\cdots,k$, which are all simultaneously applied in order to generate the desired $W$ at time $T$.
This is the method of optimal control theory. Typically, uniform finite generation requires 
a greater total evolution time $T$ than OCT methods based on pulse shaping \cite{DAlessandro}.

OCT has been applied to unitary transformation control for the purposes of quantum computation in a variety of quantum systems. Kosloff and coworkers studied the implementation of quantum gates based on vibrational eigenstates of the molecular sodium ion $Na_2^+$ on the ground electronic surface \cite{Kosloff2002,Kosloff2003}. Similar studies were carried out similar studies in the acetylene molecule using the asymmetric C-H stretching and bending modes by de Vivie-Riedle and coworkers \cite{riedle}. Gate control on spin-system dynamics with optimally designed NMR pulses has been performed by several groups \cite{Glaser1998,khaneja,matt,schirmer09}. Deutsch and coworkers implemented unitary maps on the magnetic sublevels of the ground electronic state of cesium \cite{deutsch}. The computational studies using OCT can be extended to the laboratory by shaping ultrafast laser fields using Optimal Control Experiment (OCE) \cite{judson} to generate control functions $\e(t)$.

An important issue in determining the feasibility of optimally constructing unitary transformations is how the required search effort scales with the Hilbert space dimension of the system under control. Whereas state control and maximization of observable expectation values have met with widespread success in experimental and computational incarnations \cite{constantin,ShaBru2006}, with search effort generally invariant to the Hilbert space dimension \cite{me,me2}, the achievement of high fidelity unitary transformations has proved more challenging \cite{Kosloff2002,Kosloff2003,me}, especially for large systems. 

Recently, a series of fundamental studies have been carried out on the underlying properties of quantum control landscapes, defined as the map between the external control field and the objective fidelity \cite{Glaser1998,mike1,Koch1998,oldunitary, Raj2007}. These studies revealed that under reasonable assumptions and in the absence of auxiliary costs (e.g., on the field fluence) or constraints, the control landscape contains no sub-optimal extrema, or ``traps'' that can hinder a gradient-based local algorithm for finding an optimal control field. The importance of the landscape topology to determining the feasibility of quantum control is beginning to be more widely recognized \cite{deutsch}. The topology of quantum control landscapes is dominated by so-called kinematic extremals, which are determined by the cost function alone and independent of the Hamiltonian used \cite{mike1,oldunitary,Raj2007}. Although the trap-free landscape topology ensures convergence of unconstrained gradient-based algorithms given sufficient effort, the topology does not specify the convergence rate of such algorithms, which additionally depends on the local landscape structure (e.g., slope and curvature). In this paper we examine the effects of control landscape features on the convergence rate of first-order algorithms for the optimal generation of unitary transformations. The primary goals are (i) to assign different Hamiltonians to classes exhibiting exponential or sub-exponential scaling with the system size, and (ii) to quantify the effects of the local landscape structure on the convergence rate.

The paper is organized as follows. In Section \ref{theory}, the theoretical formulation of the control problem is presented, including a summary of the associated landscape topology. Section 3 provides a framework that unifies first-order OCT algorithms for gate control, demonstrating that the convergence of all these algorithms is governed by the same underlying landscape topology. Section 4 presents metrics for landscape slope and curvature, including kinematic bounds, and dynamical metrics that quantify the effect of the control system Hamiltonian. Section 5 defines the control systems and target propagators used in the simulations. Section 6 presents numerical results on control optimization search effort and resource scaling with respect to Hilbert space dimension and identifies classes of Hamiltonians exhibiting exponential and sub-exponential scaling, while Section 7 relates search effort and resource scaling to local landscape structure. Finally, in Section 8, we draw conclusions from the findings.

\section{Optimal Control Theory}\label{theory}
\subsection{Dynamical Formulation of the Control Objective}\label{formulation}
Consider an $N$-dimensional isolated quantum system whose dynamics are governed by the time-dependent Schr{\"o}dinger equation,
\begin{equation}
i\hbar\frac{\partial U(t)}{\partial t}=H(\kappa,t)U(t),\qquad U(0,0)\equiv \mathbb{I},\label{sgl}
\end{equation}
where $H(\kappa,t)$ is the time-dependent Hamiltonian whose control variables are denoted as $\kappa$. In atomic units, the unitary propagator at some final time $T$ is
\begin{equation}
 U(T)={\bf T}{\rm exp}\left(-i\int_0^TH(\kappa,t)dt\right),
\end{equation}
where {\bf T} is the time-ordering operator and $U(T)$ is implicitly understood to be a function of $\kappa$, which in this work is represented by an external control field $\kappa\to\e(t)$. We consider an isolated quantum system satisfying the dipole formulation $H(\kappa,t)=H_0-\mu\e(t)$ where $H_0$ is the field-free (drift) Hamiltonian and $\mu$ is the dipole or control Hamiltonian operator.

This work concerns the class of control objective functionals
\begin{equation}\label{cost}
J[\varepsilon(\cdot)] = F(U(T)).
\end{equation}
The endpoint control objective $F$ may be defined as guiding the system's unitary propagator $U$ to match a pre-specified unitary matrix $W$. A convenient cost function is to minimize the Hilbert-Schmidt distance $F(U)=||W-U||^{2}$,
\begin{align}
F(U)&=||W-U||^2\nonumber\\
&=\textup{Tr}[(W-U)^{\dag}(W-U)]\nonumber\\
&={2}{N}-{2}\Re\textup{Tr}(W^{\dagger}U), \label{Jdef}
\end{align}
where the desired minimum\footnote{The minimum of $F$ corresponds to gate fidelity $\frac{1}{N}\Re\textup{Tr}(W^{\dag} U)=1$.}$F=0$ is achieved when $U=W$, and the global maximum $F=4N$ corresponds to $U=-W$. The quantum system is assumed to be controllable such that any desired $U$ can be generated by some choice of $\e(t)$ at time $T$. 
(see Section \ref{dysonsect}).

In this work, the objective $F$ of Eq. (\ref{Jdef}) is optimized using dynamical controls present in an external electric field $\e(t)$. We consider a controllable quantum system with $N$ levels $|1\rangle,\ldots,|N\rangle$ in $H_0$. To determine an optimal control $\e(t)$ that maximizes or minimizes Eq. (\ref{cost}), it is useful to define a Lagrangian functional $\bar J$ that directly imposes the dynamical constraint in Eq. (\ref{sgl}):
\begin{align}
\bar J&={2}{N}-{2}\textup{Re Tr}(W^{\dagger}U(T))\thinspace + \int_0^T \tr\left[\phi^{\dag}(t)\left(-i(H_0-\mu\e(t))U(t)-\frac{dU(t)}{dt}\right)\right]~dt \label{jsgl}\\
&={2}{N}-{2}\textup{Re Tr}(W^{\dagger}U(T))\thinspace-i\tr(\phi^{\dag}(T)U(T)) + i\tr(\phi^{\dag}(0)U(0)) +\nonumber\\ &~~+\int_0^T \textbf{H}(U(t),\phi(t),\e(t))+\tr\left(\frac{d\phi^{\dag}(t)}{dt}U(t)\right)~dt \nonumber,
\end{align}
where $\phi(t)$ is a Lagrange multiplier matrix function and $\phi(T)=U(T)U^{\dag}(t)\phi(t)$. Denoting by $\langle A,B\rangle$ the Hilbert-Schmidt inner product $\tr(A^{\dag}B)$, the first integrand term
$${\mathbf H}=- \langle U^{\dag}(T)\phi(T),iU^{\dag}(t)H_0U(t)\rangle + \e(t)\langle U^{\dag}(T)\phi(T),iU^{\dag}(t)\mu U(t)\rangle$$ is the \emph{PMP (Pontryagin maximum principle) Hamiltonian function} \cite{Jurdjevic1997,Bryson1975}. A necessary condition for maximizing or minimizing Eq. (\ref{cost}) subject to the dynamical constraint is satisfaction of the first-order conditions (Euler-Lagrange equations) for the Lagrangian $\bar J$ \cite{Bryson1975}. The first Euler-Lagrange equation is simply the \sd equation (\ref{sgl}).
The second Euler-Lagrange equation of (\ref{jsgl}) is
\begin{equation}\label{EL2}
\frac{d{\phi(t)}}{dt}=-i(H_0-\mu\e(t))\phi(t),
\end{equation}
where $\phi(T)$ satisfies the boundary condition $\phi(T)=\nabla_{U(T)}F(U(T))$. For $F(U)$ given by Eq. (\ref{Jdef}), we have \cite{jason}
\begin{equation}\label{kingrad}
\nabla_{U(T)}F(U(T,0))=U(T)W^{\dag}U(T)-W.
\end{equation}
The third Euler-Lagrange equation (critical condition) is $\frac{\partial \textbf{H}}{\partial \e(t)}=0$.
For a control system satisfying Eq. (\ref{sgl}),
\begin{align}
\frac{\partial \textbf{H}}{\partial \e(t)}&=
-i\tr\left(U^{\dag}(T)\phi(T)U^{\dag}(t)\mu U(t)\right)\nonumber\\
&=-i\tr\left(\left[W^{\dag}U(T)-U^{\dag}(T)W\right]U^{\dag}(t)\mu U(t)\right)=0.\label{EL3}
\end{align}
Consider the control-propagator map $V_T:~\e(\cdot) \mapsto U(T)$ and the composition of maps $\tilde J \equiv F \circ V_T$. Then, the functional derivative $\frac{\delta \tilde J}{\delta \e(\cdot)}$ evaluated at time $t$ is denoted as $\frac{\delta \tilde J}{\delta \e(t)}= \frac{\partial \textbf{H}}{\partial \e(t)}$. For simplicity of exposition, we use the symbols $J,\e(t)$ interchangeably with $\tilde J,\e(\cdot)$, respectively, and refer simply to the derivative $\frac{\delta J}{\delta \e(t)}$. Control fields that satisfy Eqs. (\ref{sgl}), (\ref{EL2}), and (\ref{EL3}) constitute the critical points of the control landscape $J(\e)$. Since the problem $\underset{\e(t)}{\min}~ J$  with J given by (\ref{cost}) is underdetermined, these critical points lie on \emph{critical submanifolds}, each consisting of an infinite number of fields which produce the same value of $J$.

\subsection{Critical Topology of $J$}\label{kinematics}

The matrix $-iU(T)U^{\dag}(t)\mu U(t)$ which appears in equation (\ref{EL3}) is the functional derivative $\frac{\delta U(T)}{\delta \e(t)}$. A simplifying condition that facilitates extraction of the critical topology of $J(\e(\cdot))$ is that the $N^2 \times N^2$ Hermitian Gramian matrix 
\begin{align}\label{gramian}
\mathrm{G}=\int_0^T \nu\left[U(T)\mu(t)\right] \nu^T\left[\mu(t)U^{\dag}(T)\right]~dt,
\end{align}
is full rank \cite{RajWu2008}. Here, $\nu$ denotes the ``vectorization" of an $N \times N$ complex matrix into an $N^2$-component complex vector and $\mu(t)=-iU^{\dag}(t)\mu U(t)$. Note $\mathrm{G}_{ij,kl}= \int_0^T \langle i | U(T)\mu(t) | j \rangle\langle k | \mu(t) U^{\dag}(T) | l \rangle~dt$. Satisfaction of the full-rank condition ensures that (\ref{EL3}), which may be written $\langle \nabla F(U(T)), \frac{\delta U(T)}{\delta \e(t)}\rangle=0$, implies $\nabla F(U(T))=0$ (see Section \ref{curvaturemetrics}). The condition is verified numerically for diverse classes of quantum control systems in Section 7.

When this condition is satisfied, the critical topology (number of local optima and their optimality status) of the control landscape $J(\e(t))$ is equivalent to that of $F(U)$ in Eq. (\ref{Jdef}) \cite{oldunitary,mikeu}. The control-propagator map $V_T$ associates with each critical submanifold of $F(U)$ a critical submanifold of
$J(\e(\cdot))$ whose number of positive and negative Hessian eigenvalues is identical, but which has an infinite-dimensional nullspace \cite{Raj2007}. The topology may thus be analyzed by considering the $kinematic$ degrees of freedom, using for example, the $N^2$ matrix elements of $U$ as controls.
It has been shown that there are $N+1$ distinct critical values of $F={2}{N}-{2}\textup{Re Tr}(W^{\dagger}{\hat U})$, $F$=0, 4, 8, ... $4N$ corresponding to critical points $\hat U$ where $\nabla F(\hat U)=0$ \cite{oldunitary,mikeu}.
The topology of these critical points can be determined by considering the Hessian operator ${\cal H}$ of $F$,
\begin{equation}
 {\cal H}_{ij}(W^{\dag}\hat{U})=\frac{\partial^2 F}{\partial x_i \partial x_j},
\end{equation}
where the $\{x_i\}$ are a suitable set of local kinematic coordinates around the critical point $\hat{U}$. Of interest is the number of the positive, negative, and zero eigenvalues of the Hessian at $\hat{U}$, which correspond to the number of upward, downward, and flat directions at that point.

The Hessian eigenvalue enumeration may be obtained from the Hessian quadratic form (HQF) ${\cal Q}$ of $F$:
\begin{equation}\label{hqf}
{\cal Q}_{A}(W^{\dag}\hat U)=4{\rm Re Tr}(A^2W^{\dag}\hat U),
\end{equation}
where $A$ is an infinitesimal Hermitian matrix \cite{mikeu}. At a critical point, it may be shown \cite{mikeu} that
\begin{equation}
{W^{\dag}\hat U}=R\sum_{k=1}^{N} \delta_k|k\rangle\langle k|R^{\dag},\label{v}
\end{equation}
for some unitary $R$ and where $\delta_i=\pm1$. Evaluating the HQF explicitly at a critical point and writing the elements of $A$ as $A_{ij}\equiv\alpha_{ij}+i\beta_{ij}$
yields $N^{2}$ terms ($N$ terms in the first sum and $N^2-N$ terms in the second sum) corresponding to the $N^2$ eigenvalues of the Hessian \cite{mikeu},
\begin{equation}
{\cal Q}_{A}(W^{\dag}\hat U)=4\left[\sum_{j=1}^{N}\alpha_{jj}^2\delta_j+\sum_{1\leq k<\ell\leq N}(\alpha_{kl}^2+\beta_{k\ell}^2)(\delta_k+\delta \ell)\right].\label{hqf1}
\end{equation}
The sign of each term in Eq. (\ref{hqf1}) corresponds to the sign of each Hessian eigenvalue. It has been shown \cite{mikeu,thesis} that for a critical point with an objective value of $F=4m$ for $m=0,1,2,\ldots,N$, the number of positive ($h_+$), negative ($h_-$) and zero ($h_0$) type of eigenvalue is
\begin{align}
&h_{+}=(N-m)^2;~~h_{-}=m^2;~~h_0=2Nm-2m^2. \label{hp}
\end{align}

For $F$=0, $m$=0, there are $N^2$ positive eigenvalues, indicating that the optimum is an isolated point. Similarly, for $F$=4$N$, where $m=N$, there are $N^2$ negative eigenvalues. For intermediate values of $F$, there are a mixture of positive, negative, and zero eigenvalues, indicating that all intermediate critical points have a saddle topology.
For example, at $F$=4, $m$=1, and $h_{+}=(N-4)^2$, $h_{-}=1$, and $h_0=2N-2$. Assuming that minimization of $F$ is desired, this saddle point may be expected to pose a hindrance to search effort, as there is only one direction out of $N^2$ leading down to the global minimum. The higher saddles contain more negative eigenvalues and thus are expected to pose less of a hindrance in search effort. This matter will be examined in Section \ref{saddles}.

\section{Optimization Methods}\label{gradalgs}

For unitary transformation control, deterministic first-order algorithms are typically used for control optimization. In this Section we compare these first-order algorithms and demonstrate that they share a common fixed point topology. In Section 4 we extend these results to demonstrate that the algorithms share common bounds on their convergence rates to identify optimal controls.

The simplest first-order algorithm is the gradient flow of the objective function. Using the variable $s$ to index the search path, the gradient flow trajectory is the solution $\varepsilon(s,t)$ to the initial value problem
\begin{equation}\label{Egrad}
\frac{\partial \varepsilon(s, t)}{\partial s}=
\alpha(s)~\frac{\delta J(\e)}{\delta {\e(s,t)}},
\end{equation}
for a specified initial guess for the control $\e(0,t)$, where $\alpha(s)$ is an adaptive step size. Associated with the control field trajectory $\varepsilon(s,t)$ is a trajectory for $U(s,T)$ in $\U(N)$ for the final dynamical propagator, which is induced by the control-propagator map $V_T$. In the numerical simulations in this work, Eq. (\ref{Egrad}) will be solved using a variable step size fourth order Runge-Kutta integrator built into MATLAB \cite{matlab}. A primary concern in this paper is the convergence rate of such algorithms, whose fixed points include all $\e(t)$ such that $U(T) = \hat U$ in Eq. (\ref{v}). 

As $s \rightarrow \infty$, $\e(s,t)$ converges toward stable fixed points $\bar \e(t)$ that are critical points of $J$. These points are \emph{neutrally stable}, i.e., within any neighborhood $N$ of $\bar \e(t)$ consisting of controls $\e(t)$ such that $$\norm{\e(t)-\bar\e(t)} \leq \epsilon,$$ there exists a subneighborhood $N' \subset N$ such that if $\e(0,t) \in N'$, $\e(s,t) \in N$ for all $s$.

The neutrally stable $\bar \e(t)$ solutions are the global optima of $J$, which can be seen from the corresponding trajectory $U(s,T)$ induced by the map $V_T$. $U(s,T)$ converges to \emph{asymptotically stable} fixed points that are optima of $F(U)$ - points $\hat U$ such that
$$\norm{U(0,T) - \hat U} <\delta \Rightarrow \underset{s\rightarrow \infty}{\lim} U(s,T) = \hat U,$$
for some $\delta$ that is equal to the radius of the attracting region of the fixed point. The latter are the critical points identified in Eq. (\ref{v}) with positive definite Hessian (\ref{hqf1}) - and according to (\ref{hp}), the only critical point satisfying this criterion is the unique global optimum $W$. Due to the asymptotic stability of $\hat U=W$,
any $\e(0,t)$ such that $U(0,T)=V_T(\e(0,t))$ is within the attracting region of $W$ 
will converge to a neutrally stable fixed point $\bar \e(t)$ that lies on
the global optimum submanifold of $J$. The \emph{instability} of fixed points $\hat \e(t)$ lying on
other critical submanifolds with $J > 0$ follows from the indefiniteness of the HQF at $\hat U \neq W$.

Many unitary control studies use so-called \emph{PMP-iterative} algorithms \cite{turinici}, which can be formulated only in discrete time. These algorithms iteratively integrate equations (\ref{EL2})  and (\ref{sgl}) at each step $k$, using control fields $\tilde \e_k(t) = \alpha_k \e_{k-1}(t) + \beta_k \langle \phi_k(t) | \mu | U_{k-1}(t)\rangle$,
$\e_k(t) = \alpha_k\tilde \e_k(t) + \beta_k\langle \phi_k(t) | \mu | U_k(t) \rangle$, respectively, where $\alpha,\beta$ are scalars. The fixed points of these algorithms are points on the control landscape where $\e_k-\tilde \e_k=0$ or $\tilde \e_k-\e_{k-1}=0$. In Appendix \ref{pmpiter} we show that under appropriate regularity conditions the only neutrally stable $\bar \e(t)$ lie on the global optimum submanifold where $U(T)=W$. A third class of gate control optimization algorithms consists of first-order \emph{tracking} algorithms which follow a prescribed path in the space of propagators to the target gate; these have been shown to be capable of achieving gate fidelities approaching machine precision \cite{jason}. In this work we focus on the application of steepest descent algorithms (\ref{Egrad}), due to the mathematical convenience of formulating convergence to a stable fixed point for these continuous time algorithms, but our conclusions on convergence efficiency 
are applicable to PMP-iterative and tracking algorithms as well.

\section{Landscape Structure Metrics}\label{str}
The landscape topology summarized in Section \ref{kinematics} suggests that an optimal field to achieve a desired unitary transformation may be readily found because no suboptimal extrema exist on the landscape. This attractive behavior does not preclude the possibility that complicated landscape features, including strong influence by saddle regions, may impede optimal searches. Thus, an understanding of the effects of the local landscape structure on optimal searches is necessary in order to explain and predict the scaling of search effort with system size $N$. We introduce landsape metrics and show that the same landscape local structural features govern the convergence of PMP-iterative and gradient-based algorithms.

\subsection{First-order metrics}

The local structure metrics of the landscape are based on a Taylor expansion $J(\e(s,t) + \delta \e(s,t)) = J(\e(s,t)) + \int_0^T \nabla_{\e}J(\e)\delta\e(s,t)~dt + \frac{1}{2}\int_0^T \int_0^T \mathcal{H}(t,t')\delta\e(s,t)\delta\e(s,t')~dtdt'+\cdots$ of the cost functional $J$ with respect to $\e(s,t)$.
The slope metric $\mathcal{G}_m$ at a point $s_m$ on the landscape is defined as
\begin{equation}
\mathcal{G}_m=\nabla J\bigr\rvert_m\cdot \overrightarrow{u_m}\equiv||\nabla J\bigr\rvert_{m}||= \left(\int_0^T dt\left(\frac{\delta J}{\delta\e(s_m,t)}\right)^2\right)^{1/2}.\label{steep}
\end{equation}
where the unit gradient vector is $\overrightarrow{u_m}\equiv \frac{\overrightarrow{\nabla} J\rvert_{m}}{||\nabla J\rvert_{m}||}$. The metric \gr$_m$ is thus equivalent to the magnitude of the gradient on the landscape at the $m$th point. Beginning from the expression in Eq. \ref{steep}, ${\cal G}_m$ at any point $m$ is bounded by
\begin{align}
\left|\nabla J(\varepsilon(t))\right|&=
\left(\int_0^Tdt\left|\tr\left[(W^{\dag}U(T)-U(T)^{\dag}W)\mu(t)\right]\right|^2\right)^{1/2}\nonumber\\
&\leq\left(\int_0^Tdt\left(|\langle W^{\dag}U(T), \mu(t) \rangle + \langle U^{\dag}(T)W\mu(t) \rangle|\right)^2\right)^{1/2}\nonumber\\
&\leq\left(\int_0^Tdt\left(||W^{\dag}U(T)||||\mu(t)||+ ||U^{\dag}(T)W||||\mu(t)||\right)^2\right)^{1/2}\nonumber\\
&=2N\sqrt{T}||\mu||\label{normg}.
\end{align}
Above, the Cauchy-Schwarz inequality is used twice. A greater value of \gr$_m$ at $m=0,1,2,\ldots$ results in a locally faster descent.

We can also establish a bound on $|\e_k(t) - \tilde \e_k(t)|$ in PMP-iterative algorithms (see Appendix \ref{pmpiter}):
\begin{align*}
|\langle \phi_k(t) | \mu | U_{k-1}(t) \rangle| &= |\tr[U_k^{\dag}(T) (W - U_{k-1}(T) W^{\dag} U_{k-1}(T))U_k^{\dag}(T)),U_{k-1}^{\dag}(t)\mu U_k(t)\rangle]|\\
&\leq \norm{U_k(t) U_k^{\dag}(T) W U_{k-1}(t)}\norm{\mu} + \norm{U_k(t)U_k^{\dag}(T)U_{k-1}(T)W^{\dag}U_{k-1}(T)U_{k-1}(t)}\norm{\mu}\\
&\leq 2N\norm{\mu}.
\end{align*}
Thus,
\begin{align*}
\norm{\e^{(k)}(t) - \bar \e^{(k)}(t)} \leq 2N\sqrt{T}\norm{\mu}.
\end{align*}
In PMP-iterative algorithms, the bound on the increment in the field for infinitesimally small step length is equivalent to that for steepest descent with a Mayer cost.

\subsection{Second-order metrics}\label{curvaturemetrics}

From the second variation of the objective functional $J$, we may derive the Hessian kernel; the elements of the Hessian are given by \cite{jason}
\begin{equation}\label{hessiankernel}
{\cal H}(t,t')=\tr\left[W^{\dag}U(T)\mu(t)\mu(t')+U^{\dag}(T)W\mu(t')\mu(t)+(W^{\dag}U(T)-U^{\dag}(T)W)[\mu(t),\mu(t')]\right].
\end{equation}At a critical point, the last term of Eq. (\ref{hessiankernel}) drops out. The relationship between the HQF expression (\ref{hqf}) and (\ref{hessiankernel}) is described in Appendix \ref{hess}.

In steepest descent algorithms, according to the gradient expression (\ref{EL3}),
$\delta \e(t)$ is composed of linear combinations of the real and imaginary
components of $\mu_{ij}(t)$. The eigenvalues of the Hessian (\ref{hessiankernel}) specify the rates at which new frequency modes required for optimal control (contained within the eigenfunctions of $\mathcal{H}(t,t')$) can be added to $\delta \e(t)$. Thus, several bounds on the Hessian are given below.

The Hessian trace or mean curvature is given by
\begin{equation}
{\rm Tr}{\cal H}(t,t')=\int_0^T dt{\cal H}(t,t)\label{tr}.
\end{equation}
At a critical point,
\begin{align}
{\rm Tr}{\cal H}(t,t')&=\int_0^Tdt{\rm Tr}\left[W^{\dag}U(T) U^{\dag}(t)\mu U(t)U^{\dag}(t)\mu U(t)+U^{\dag}(T)W U^{\dag}(t)\mu U(t)U^{\dag}(t)\mu U(t)\right]\nonumber\\
&=\int_0^Tdt{\rm Tr}\left[W^{\dag}U(T) U^{\dag}(t)\mu^2 U(t)+U(T)W^{\dag}U^{\dag}(t)\mu^2 U(t)\right]\label{trc}
\end{align}
At $J$=0, $W^{\dag}U(T)=I_{N}$, and Eq. (\ref{trc}) becomes
\begin{align}
{\rm Tr}{\cal H}(t,t')\bigr\rvert_{J=0}&=2\int_0^Tdt{\rm Tr}\left[U^{\dag}(t)\mu^2 U(t)\right]\nonumber\\
&=2\int_0^Tdt{\rm Tr}\left[U^{\dag}(t)U(t)\mu^{2}\right]\nonumber\\
&=2T {\rm Tr}\mu^2\label{tro}
\end{align}
where the second step uses the cyclic permutation trace rule. Similarly, at the maximum $J$=4$N$, the trace is given by the negative value of Eq. (\ref{tro}).

We can also calculate a bound on Hessian mean curvature away from a critical point:
\begin{align*}
\tr(\mathcal{H}(t,t')) &= \int_0^Tdt\tr\left[W^{\dag}U(T) U^{\dag}(t) \mu^2 U(t) + U(T)W^{\dag}U(t)\mu^2 U(t)\right]\\
&=\int_0^Tdt\left(\norm{W^{\dag}U(T)}\norm{\mu(t)^2} + \norm{U^{\dag}(T)W}\norm{\mu(t)^2}\right) \\
&= 2NT\norm{\mu^2}.
\end{align*}

Finally, we consider the local curvature, or the projection of the Hessian matrix on to the normalized gradient vector $\overrightarrow{u_m}$,
\begin{equation}\label{localcurv}
 {\cal C}_m=\overrightarrow{u_m}\cdot {\cal H} \cdot \overrightarrow{u_m}'.
\end{equation}
The curvature near the optimum may influence the required search effort by determining the ease of convergence to the optimum.
Note that since the gradient and Hessian can be expressed in terms of the same $N^2$ basis functions of time,
only $\mathcal{L}^2$ inner products of components of the time-evolved dipole operator contribute to the local curvature of the control landscape.

\subsection{Distance metrics}
On a search trajectory, the field starts out at algorithmic index $s$=0 with $\e(0,t)$ and progresses in steps $s\to s+ds$ (i.e., $\e(s,t)\to\e(s+ds,t)$) until the trajectory ends at an optimal field, $\e_{opt}=\e(s_{M},t)$ at $s=s_{M}$. The complexity of the search may be characterized by the ratio of the trajectory path length $||\Delta_P\e(t)||$ to the Euclidian distance between initial and final control fields $||\Delta_E\e(t)||$,
\begin{align}
R_{\e}&=\frac{||\Delta_P\e(t)||}{||\Delta_E\e(t)||}=\frac{\int_0^{s_{M}}ds\left(\int_0^T dt \left[\frac{d\e(s,t)}{ds}\right]^2\right)^{1/2}}{\left(\int_0^T dt \left[\e(s_{M},t)-\e(0,t)\right]^2\right)^{1/2}}\label{ratio}
\end{align}
The closer $R_{\e}$ is to unity, then the more direct the path, i.e., the closer the path is to a straight line in search space. This metric will be used to assess the complexity of the search trajectories followed during optimizations.

Since the presence of saddle manifolds on the landscape may influence the efficiency of an optimal search, the distance of points on the search trajectory to the nearest saddle also provides important structural information. This distance may be measured by examining the eigenvalues of the matrix $V=W^{\dag}U$, since these eigenvalues are all $\pm 1$ at any saddle (all $+1$ at $J$=0 and all $-1$ at $J$=$4N$). If $E_i$ are the eigenvalues of $V$, a convenient metric to express the distance to the nearest saddle is
\begin{equation}
 {\cal S}={\cal N}\sum_{i=1}^N (1-|{\rm Re} E_i|),\label{smetric}
\end{equation}
where the normalization factor $\cal{N}$ is ${\cal N}=2/J$ if $J\leq2N$ and ${\cal N}=2/(4N-J)$ if $J>2N$, which makes the maximal allowed value of ${\cal S}$ equal to 1 for any $J$ value. Finding that ${\cal S}$ is close to zero near the saddle values ($J$=4, 8, 12, ...) indicates that the search trajectory encounters the saddle manifold at this $J$ value. The effects of search trajectories approaching saddle manifolds will be examined in detail in Section \ref{saddles}.

\subsection{Gramian matrix}\label{gramsection}

Unlike the Hessian, which depends on $F(U)$ as well as the system
Hamiltonian, the Gramian matrix $\mathrm{G}_{\e,T}$ (\ref{gramian}) provides a means of
characterizing purely dynamical effects on the optimization trajectory. Consider the equation for the controlled
propagator $U(s,T)$ corresponding to the gradient flow (\ref{Egrad}).
Denoting by $u(s,T)$ the vectorization of the propagator $U(s,T)$, we have \cite{RajWu2008}:
\begin{equation}\label{gramflow}
\frac{\partial u(s,T)}{\partial s} = \mathrm{G}_{\e_s,T}\nabla_u F(u(s,T)), \end{equation}
from which it can be seen that $\mathrm{G}_{\e,T}$ is a linear map between vectors in $T_U U(N)$ and all system-dependent effects.
The eigenvectors of the Gramian are $N^2$ orthogonal directions in the tangent space to the unitary group $\mathcal{T}_U U(N)$. The magnitudes of the eigenvalues of $\mathrm{G}_{\e,T}$
represent the dynamical contributions of the corresponding orthogonal directions in $\mathcal{T}_U U(N)$
to the propagator variation $\delta U(s,T)$ induced by the gradient flow control variation $\delta \e(s,t)$.

If the eigenvalues of $\mathrm{G}_{\e,T}$ are always sufficiently far from zero, corresponding to
a well-conditioned matrix, then
$\norm{\frac{\delta J}{\delta \e(t)}}$ can be infinitesimally small only near the global optimum submanifold,
irrespective of the direction of $\nabla_U F(U(T))$,
which facilitates convergence. This can be seen as follows: the critical point
condition $\nabla J=0$ implies
$$\big\langle \nabla_U F(U(T)), \frac{\delta U(T)}{\delta \e(t)}\big\rangle =0,~\forall t;$$
then if $\nabla_UF(U)\neq 0$ at a critical point, then $\mathrm{rank}~\mathrm{G}_{\e,T} < N^2$. Since the only
neutrally stable fixed points of flow (\ref{Egrad}) lie on the global optimum manifold, the claim holds.

More generally, control systems for which the expected values of the condition number of $\mathrm{G}_{\e,T}$ are low near the global optima typically exhibit faster convergence, as shown in Section 7. A special case is where $\mathrm{G}_{\e,T}$ is degenerate; then, the dynamical contribution of each eigenvector in $T_U U(N)$ contributes equally when the control variation is integrated over time and convergence is governed by the kinematic gradient\footnote{In prior work on gate control optimization, the target gate $W$ was sometimes chosen to reside in a subspace of the dynamical Hilbert space \cite{Kosloff2002,Kosloff2003}. In this case, operators on the Hilbert subspace need not be unitary (rather, they belong to the class of Kraus operators or positive trace-preserving maps), and Hamiltonian-dependent contributions to optimization efficiency are governed primarily by the eigenvalues of the Gramian matrix (\ref{gramian}) on that subspace rather than the entire Hilbert space.}.

\subsection{Higher-order analysis}\label{dysonsect}

The above metrics and associated bounds characterize the local properties of the control landscape.
Properties of the \emph{global} solution to the gate control problem
can be analyzed using the Dyson series expansion for the controlled unitary propagator in the interaction picture \cite{abhra}:

\begin{align}\label{dysonexp}
U_I(T) &= I_N + i\int_0^T V^{\dag}(t^1) \mu V(t^1)\e(t^1)~dt^1 +  \int_0^{T} V^{\dag}(t^1) \mu V(t^1)\e(t^1) \int_0^{t^1} V^{\dag}(t^2) \mu V(t^2)\e(t^2)~dt^2dt^1+\cdots 
\end{align}
where $V(t) = \exp(-iH_0t)$. The $n$-th term in the Dyson expansion corresponds physically to the set of possible $n$-photon transition pathways between eigenstates over time $[0,T]$. Note that each successive term in (\ref{dysonexp}) contains higher order products of $\exp(iH_0t)\mu \exp(-iH_0t)$ and $\e(t)$ than the previous terms. For any bounded field fluence $\int_0^T \e^2(t)~dt$ and tolerance $c$, the series converges to $U_I(T)$ at some finite order \cite{abhra}.

The matrices $H_0$ and $\mu$, which define the control system,
also fully determine the \emph{minimal} order in series (\ref{dysonexp}) required
to produce any given $W$. Let $\mathcal{L}\{H_0,\mu\}$ denote the \emph{dynamical Lie algebra} of the quantum control system - i.e., the Lie algebra spanned by repeated commutators of $H_0$ and $\mu$:
$$\mathcal{L}\{H_0,\mu\}=\mathrm{span}~\{H_{i_k},\cdots,\{H_{i_2}, H_{i_1}\}\},~H_i \in \{H_0,\mu\}.$$ Beyond some critical $k$ (called the \emph{depth} of $\mathcal{L}$),
which depends on the control system, the dynamical Lie algebra saturates \cite{DAlessandro}.
For bilinear quantum control systems,
a sufficient condition for full controllability - i.e., the existence of a control field $\e(t)$ such that the corresponding $U(T)$ induced by the \sd equation can be any unitary matrix - is that $\mathrm{rank}~ \mathcal{L}\{H_0,\mu\}=N^2$ \cite{DAlessandro,ramakrishna}.

Higher-order terms in series (\ref{dysonexp}) correspond to higher order commutators, and hence generally require higher field fluences (higher amplitudes of the corresponding Fourier modes). If these amplitudes in $\e(t)$ are below certain minimal values required for the corresponding Dyson series terms to be nonnegligible, it is not possible for the field to be a solution to the gate control problem. In general, \emph{systems with control Hamiltonian (dipole) operators $\mu$ that systematically exclude distant transitions require higher order terms in the Dyson series} to reach any arbitrary gate $W$, which increases the nonlinearity of the optimization problem for these control systems with greater Lie algebra depth $k$. This results in the optimal controls $\bar \e(t)$ containing more frequency modes, corresponding to more complex control mechanisms \cite{abhra}. In Sections 6 and 7, we demonstrate that optimizing with control Hamiltonians with weak or forbidden transitions between distant quantum states yields higher-fluence and more complex optimal controls $\bar \e(t)$, which require a greater algorithmic search effort to find.

Equation (\ref{dysonexp}) also determines when the Gramian matrix (\ref{gramian}) may be nonsingular. Only for controls $\e(t)$ where all the terms in the Dyson series required for full controllability of a given $H_0$ and $\mu$ have become populated can the Gramian matrix be well-conditioned.
This is quantified by the relation\footnote{We prove this result, as well as other necessary conditions for nonsingularity of the Gramian, in a separate work \cite{raj3}.}
$$\mathrm{rank}~\mathrm{G}_{\e,T}\leq \mathrm{rank}~ \mathcal{L}\{H_0,\mu\}.$$ In Section 7, the effects of control system Hamiltonians on Gramian condition number are studied numerically.

\section{Control systems}\label{physical}
An exhaustive sampling of system structures for the drift Hamiltonian $H_0$ and the control Hamiltonian $\mu$ is impractical. Here, we choose two $H_0$ and four $\mu$ structures motivated by common propagator control systems, which differ qualitatively in their Lie algebra depth, as described in Section 4.E. The goal is to provide an overview of search behavior that might be expected without making any specific predictions for the behavior of any particular quantum system. In order to delimit the space of drift and control Hamiltonians studied, only diagonal $H_0$ structures are considered, with structural variation restricted to the control Hamiltonians $\mu$. Although control systems requiring multiple fields to ensure controllability (e.g., coupled spin systems) may be used to realize quantum computation \cite{schirmer09}, we consider only systems controllable by a single field here so that the search behavior across different systems can be directly compared.

We consider an $N$-level quantum system in arbitrary dimensionless units. Two model systems with $H_0$ in its diagonal basis are considered. First is that of a rigid rotor,
\begin{equation}
H_0=\sum_{j=0}^{N-1} \frac{1}{2} \thinspace j\left(j+1\right)|j\rangle\langle j|, \label{ho}
\end{equation}
and second is that of an anharmonic oscillator,
\begin{equation}
H_0=\sum_{j=0}^{N-1}\left[\omega(j+1/2)-\frac{\omega^2}{\cal D}(j+1/2)^2\right]|j\rangle\langle j|\label{osc},
\end{equation}
with $\omega=20$ and ${\cal D}=2000$.

Four physically relevant real matrix control Hamiltonian structures $\mu$ will be considered, paired with one of the two $H_0$ structures above. Control Hamiltonians of different matrix distributions of off-diagonal elements were chosen because of their different Lie algebra depths and optimal control mechanisms, as discussed in Section \ref{dysonsect}. All of the control Hamiltonian structures used here have nonzero trace in order to make the systems controllable on $U(N)$ and not only on $SU(N)$. For many physical systems the coupling between states decreases as the difference between the quantum numbers of the states increases, and the first choice of $\mu$ takes this property into account, with the following structure
\begin{equation}
\mu=
\begin{pmatrix}\alpha& 1&D& D^2& \ldots& D^{N-2}\\
1& \alpha& 1& D& \ldots& D^{N-3}\\
D& 1& \alpha& 1& \ldots& D^{N-4}\\
D^2& D&1& \alpha&\ldots& D^{N-5}\\
\vdots& \vdots& \vdots& \vdots& \ddots& \vdots\\
D^{N-2}&D^{N-3}& D^{N-4}& D^{N-5}&\ldots& \alpha
\end{pmatrix}\label{mu}
\end{equation}
where $\alpha>0$, $D\in[0,1]$ is the coupling parameter, and all elements of $\mu$ have a random phase of $\pm 1$ with the restriction that $\mu$ remains symmetric. This ``$D$'' structure qualitatively corresponds to diatomic molecules and other anharmonic vibrational systems. The second control Hamiltonian structure examined is the related ``banded'' structure where a fixed number of rows nearest to the diagonal have elements of $\pm1$ with the remaining rows having elements of zero; the extreme of having only one row with allowed transitions qualitatively corresponds to a harmonic oscillator or rigid rotor. Third, we consider a ``sparse'' structure with 50$\%$ of the off-diagonal elements randomly chosen as $\pm1$ and the remaining 50$\%$ of the off-diagonal elements being zero, while maintaining $\mu$ as symmetric. Sparse control Hamiltonians with fewer than 50$\%$ allowed couplings are examined as well in Section \ref{scaled}. Control Hamiltonian structures containing some allowed and some forbidden transitions qualitatively correspond to coupled-spin system qubit structures commonly used in quantum computation, although only certain specified distributions of couplings are allowed for qubit systems. In order to investigate the search effort for systems with such control operators, we consider a ``tensor product'' control Hamiltonian on $n$ qubits,
\begin{equation}
 \mu=\sum_{j=1}^n \sigma_x^j+\alpha{\mathbb I},\label{qubit}
\end{equation}
where the diagonal matrix is added to make the system controllable on $U(N)$ and $\sigma_x^j$ is
$$\sigma_x^j = \underset{j-1}{\underbrace{I_2 \otimes \cdots \otimes I_2}} \otimes \sigma_x \otimes \underset{n-j}{\underbrace{I_2 \otimes \cdots \otimes I_2}}.$$
We consider pairings of this $\mu$ with the diagonal $H_0$ operators above
\footnote{Systems with multiple control fields, each associated with a
different Pauli operator (required for full controllability of coupled spins),
are considered in a separate work.}.

As we will demonstrate in Sections 6 and 7, the distribution of couplings between states is important for assessing the scaling of effort with $N$. With this in mind, comparing the sparse and tensor product structures $\mu$ reveals some important differences, for example at $N$=8.
\begin{equation}
 \mu_{tensor}=
\begin{pmatrix}\alpha& 1&1& 0& 1&0&0&0\\
1& \alpha& 0&1& 0& 1& 0&0\\
1& 0& \alpha& 1& 0&0&1&0\\
0& 1&1& \alpha&0&0&0&1\\
1&0&0&0&\alpha&1&1&0\\
0&1&0&0&1&\alpha&0&1\\
0&0&1&0&1&0&\alpha&1\\
0&0&0&1&0&1&1&\alpha
\end{pmatrix}\qquad\qquad
\mu_{sparse}=
\begin{pmatrix}\alpha&1&0&0&1&1&0&1\\
1&\alpha&1&1&1&0&0&1\\
0&1&\alpha&0&1&0&0&0\\
0&1&0&\alpha&0&1&1&1\\
1&1&1&0&\alpha&0&1&0\\
1&0&0&1&0&\alpha&0&1\\
0&0&0&1&1&0&\alpha&0\\
1&1&0&1&0&1&0&\alpha
\end{pmatrix}
\end{equation}
While each operator has an equal number of allowed transitions, the tensor product structure allows no transitions more than four states apart (note the zeros in the upper-right and lower-left corners of the matrix). The sparse $\mu$ example shown, in contrast, allows transitions between states $|1\rangle$ and $|8\rangle$, as well as some other transitions five or more states apart. Structural differences in coupling distributions are even more evident at $N$=16 and $N$=32. We thus define two distinct classes of control Hamiltonians: those that allow distant transitions, including the $D$=1.0 and sparse structures, and those that forbid or have very weak distant transitions, including $D<$1.0, banded, and tensor product structures. The differences in coupling distributions between these two classes influence the required search effort, as will be shown in Sections 6 and 7.

The simulations will consider both random Haar-distributed unitary $W$ matrices \cite{mezzadri} and the Fourier transform quantum gate,
\begin{equation}
W_{FT,n}(j,k) = \frac{1}{2^n}\exp(\frac{2\pi \imath \cdot j*k}{2^n}),\label{ft}
\end{equation}
where $j$ and $k$ denote the matrix elements and run from 1 to $N=2^n$.

The initial field at $s=0$ is chosen as
\begin{equation}\label{fieldexp}
\e(0,t)=f\textup{exp}\left[-\frac{8\pi}{T^2}\left(t-\frac{T}{2}\right)^2\right]\sum_{k=1}^{K}\textup{sin}\left(\omega_kt+\phi_k\right),\thickspace t\in[0,\thinspace T]
\end{equation}
where $\{\omega_k\}$ are the $K$ Fourier components of the field, which are selected randomly and bounded by the frequency of the $|1\rangle\to|N\rangle$ transition in $H_0$, $\phi$ is a random phase on $[0,2\pi]$, and $f^2$ is the field fluence. Prior to multiplication by $f$, the field is normalized to have unit fluence.

\section{Control Search Complexity}\label{scale}
The search effort required to find an optimal field $\e(t)$ has important implications for determining the feasibility of controlling the dynamics of complex systems. In Section \ref{saddles}, the influence of the saddle point topology of the control landscape is assessed. In Section \ref{scale1}, we examine the search effort as a function of $N$ for a broad range of choices of $H_0$, $\mu$, initial field strength $f$, and $W$. Further exploration of the control Hamiltonian structure's effect on search effort in Section \ref{scaled} identifies control Hamiltonian properties that result in the most efficient searches. Details of the numerical parameters in the simulations are given in Appendix \ref{num}.

\subsection{Influence of Landscape Saddle Point Topology}\label{saddles}

The simulations in this section address how the landscape topology, which is primarily determined by the kinematic cost function $F$, influences the behavior of gradient-based optimizations for systems of dimension up to $N$=8. In particular, we examine the extent to which the saddle manifolds influence the search trajectory and whether the saddle effects are dependent on the choice of $W$ or the initial control field. The Hamiltonian $H_0$ is given by Eq. (\ref{ho}) and $\mu$ given by Eq. (\ref{mu}) with $D$=1.0, 0.9, or 0.6.

\begin{figure}
\includegraphics[width=8.5cm]{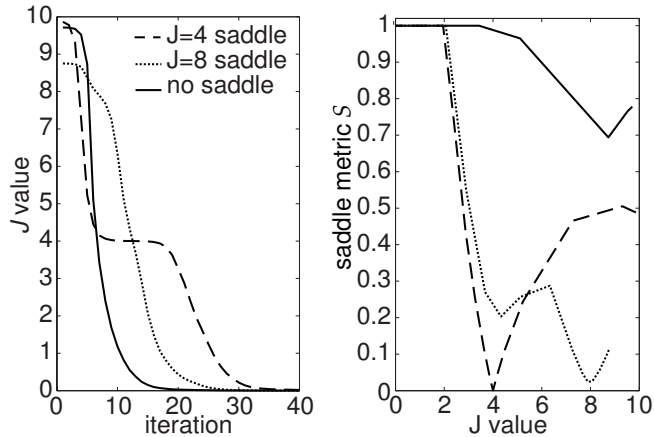}
\caption{Sample search trajectories for $N$=4. Left panel: $J$ value versus iteration. The solid-line trajectory goes directly from the initial $J$ value to $J$=0, while the dashed and dotted trajectories slow down at $J$=4 and $J$=8, respectively, suggesting interaction with the corresponding saddles. Right panel: Saddle metric versus $J$ for the same searches.\label{saddleex}}
\end{figure}

\begin{figure}
 \includegraphics[width=8.5cm]{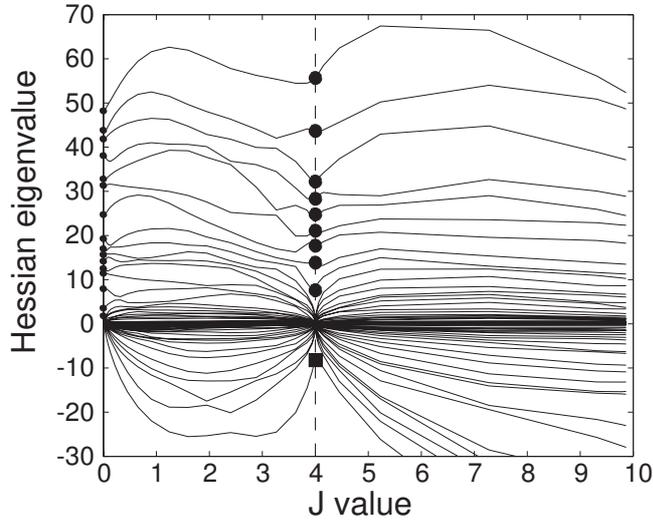}
\caption{Hessian eigenvalues versus $J$ value for the dashed-line trajectory of Figure \ref{saddleex}. This search encountered the $J$=4 saddle, shown by the nine positive Hessian eigenvalues (labeled by large circles) and one negative eigenvalue (labeled by a square). At the optimum, there are 16 positive Hessian eigenvalues (small circles). All values are unitless. \label{hesseig}}
\end{figure}

The trajectories of three searches for $N$=4 are shown in Figure \ref{saddleex}.
Comparison of the saddle metric ${\cal S}$ (c.f., Eq. \ref{smetric}) in the right panel with the optimization trajectories in the left shows that interaction with saddles retards convergence to the optimum. Examination of the trajectory of the Hessian eigenvalues during the search confirms interaction with a saddle manifold. The Hessian eigenvalues of the search interacting with the $J$=4 saddle are shown versus $J$ in Figure \ref{hesseig}. At $J$=4 (dotted vertical line), there are nine positive eigenvalues (marked by circles) and one negative eigenvalue (marked by square), in agreement with the Hessian spectrum derived in Eq. (\ref{hp}). Furthermore, at the optimum $J$=0, there are 16 positive Hessian eigenvalues (marked by small circles), in agreement with the maximally allowed $N^2$ positive eigenvalues. The remaining Hessian eigenvalues are null, as predicted \cite{oldunitary,mikeu}.

Table \ref{saddletable} presents statistics on optimizations using a variety of conditions. 1000 searches starting from different initial fields were used to generate the statistics. Shown is the required search effort (defined as the number of algorithmic iterations to reach $J<10^{-6}\times J_{max}$) as well as the fraction of searches that interact with saddles. Three degrees of saddle interaction are examined: ${\cal S}<0.1$, ${\cal S}<0.05$, and ${\cal S}<0.01$. The probability of saddle interaction decreases with rising Hilbert space dimension $N$, such that by $N$=8, negligibly few searches have strong interactions with saddles. The decrease in saddle interactions as $N$ rises is favorable to performing large-scale unitary transformation optimizations.

\subsection{Scaling of effort with $N$}\label{scale1}

Simulations were performed for $N$=2, 4, 8, 16, and 32 with a statistical sample size of 20 (with the exception for some cases of $N$=32, where a single optimization was performed) and a convergence criterion of $J<10^{-3}\times J_{max}$. The mean search effort with statistical error is shown in Table \ref{scaletable} for all optimizations. The effort is plotted versus $N$ for $D$=1.0, 0.9, sparse, and tensor product structures $\mu$ with rotor $H_0$ in Figure \ref{deffort}. The observed scaling of effort for a fixed $\mu$ structure is similar for both the rotor (Eq. (\ref{ho})) and oscillator (Eq. (\ref{osc})) $H_0$ structures, as seen in Table \ref{scaletable}. The search effort scaling was found to be strongly dependent on the control Hamiltonian structure. For $D$=1.0 or sparse $\mu$, i.e., class that allows distant transitions between states, the effort scales slowly with $N$. In contrast, for the $D<$1.0 and tensor product structure (Eq. (\ref{qubit})), i.e., the class that forbids or has weak distant transitions, the effort scales exponentially with $N$, as shown by the least-square fit lines on the semi-log plot for $D$=0.9 and tensor product $\mu$ structures in Figure \ref{deffort}.

The fluence of the initial field has some effect on the absolute search effort, but not on its scaling with $N$ (columns 4 and 5 of Table \ref{scaletable}). Increasing the fluence cannot overcome exponential scaling for the class of control Hamiltonians that forbids distant transitions. The effort can be reduced to some extent by allowing $T$ to scale with $N$ (Table \ref{scaletable}), in agreement with the conclusion that longer control times are needed for systems that have few accessible control pathways \cite{sussmann}, but the effort still scales exponentially with $N$. The choice of $W$ (i.e. random unitary or FT gate) does not greatly affect the search effort, as shown by comparing the effort to find the FT gate or a random $W$ using the sparse $\mu$ structure.

\subsection{Control Hamiltonian Structure and Search Effort}\label{scaled}
Of all the search parameters explored above, only the control Hamiltonian structure has a systematic effect on the scaling of the search effort with $N$. In order to determine the effects of the structure of $\mu$ for fixed $N$ (here $N$=8), we compared control Hamiltonians with $D$=1.0, 0.9, 0.75, and 0.6, randomly generated sparse structures with 14, 10, or 8 allowed transitions, and banded control Hamiltonian structures where 2, 3, or 4 rows nearest to the diagonal contain allowed transitions. The resulting search effort is plotted versus the norm $||\mu||$ in Figure \ref{dip}, which clearly shows that $||\mu||$ does not determine search effort. Rather, the distribution of strong couplings between states is important.

For a given value of $||\mu||$, the banded structure has the greatest search effort, followed by the $D$ structure, and the sparse structure has the smallest search effort. For approximately the same value of $||\mu||\simeq 5$, the search effort varies by over a factor of 10, from 70 iterations for the sparse structure (14 transitions), through 100 iterations for $D$=0.75 structure, to 1200 iterations for the banded structure with 2 rows of allowed transitions (13 transitions). This indicates that systematic exclusion or suppression of transitions between distant states while allowing only transitions between near states raises the search effort, compared to having an equal number of allowed transitions that include some distant transitions. Thus, the control Hamiltonian class that allows distant transitions is expected to have a lower search effort than the class that forbids distant transitions. This observation is consistent with the observed exponential scaling of the search effort with $N$ for $D<$1 and tensor product structures. The reasons behind the dependence of the search effort on the control Hamiltonian structure will be explored in Section \ref{searchstruct}.

\section{Search Effort and Landscape Geometry}\label{searchstruct}

Here, we assess the local landscape features in terms of the metrics in Section \ref{str} for unitary propagator control. We first consider the structure of the landscape in terms of the local metrics in Section \ref{local}. The effects of the landscape structure on the search trajectories, as defined by the directness metric $R_{\e}$ and the Gramian matrix, are examined in Section \ref{distance}.

\subsection{Local Landscape Structure}\label{local}
The bound on the slope metric ${\cal G}$ derived in Section \ref{str} was found to be conservative. The recorded maximal slope metric ${\cal G}$ was always significantly below this bound and observed to grow linearly with $N$, while the bound grows quadratically with $N$ (not shown). The analytically derived Hessian trace at $J$=0 (c.f. Eq. \ref{trc}) was found to hold; the deviation at the optimum was always less than 0.001$\%$ when the convergence criterion $J\leq1e-6J_{max}$ was used. The Hessian trace does not predict search effort regardless of where it is measured, since it is only dependent on $||\mu||$ or $||\mu^2||$, and the effort can vary widely for different $\mu$ structures with similar values of $||\mu||$ (c.f., Figure \ref{dip}).

\begin{figure}
 \includegraphics[width=8.5cm]{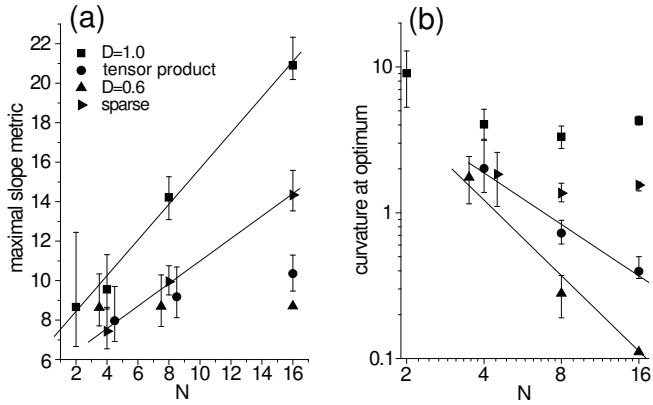}
\caption{(a) Mean value of the maximal slope metric \gr$_{max}$ versus $N$ for $D$=1.0 (squares), tensor product (circles), $D$=0.6 (up triangles), and sparse $\mu$ structure (side triangles).
(b) Mean curvature \curv\ms at the optimum versus $N$. The rotor $H_0$ structure is used for all searches. All values are unitless.\label{gc}}
\end{figure}

\begin{figure}
 \includegraphics[width=8.5cm]{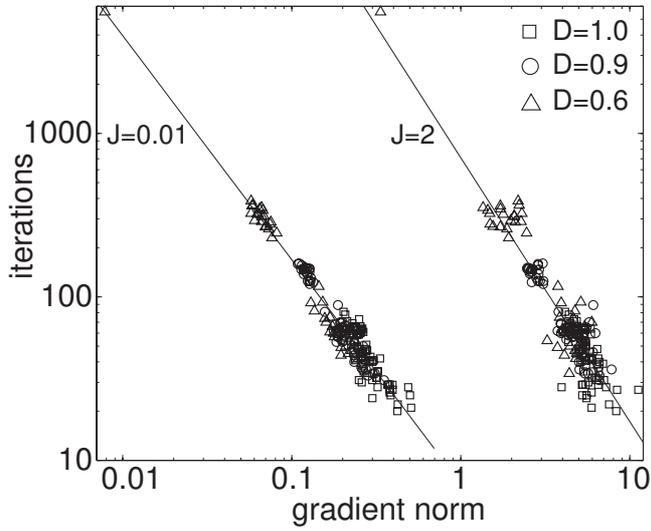}
\caption{Search effort versus slope metric measured at $J$=0.01 and $J$=2. The effort scales with slope metric according to a power law, as seen by the least squares lines on the log-log plot. All values are unitless.\label{itergrad}}
\end{figure}

The slope metric ${\cal G}$ and the local curvature ${\cal C}$ were found to correlate with search effort. The statistical distribution of the maximal slope metric ${\cal G}_{max}$ over the search samples is plotted versus $N$ in Figure \ref{gc}(a). For $D$=1.0, the maximal slope rises linearly with $N$, but the growth with $N$ is slower for $D<$1.0 and tensor product control Hamiltonians. Since the maximal slope metric for any search is often recorded at or near the initial field (depending on the exact choice of $\e(0,t)$), an estimate of search effort scaling with $N$ for any $\mu$ structure can be made simply by measuring the gradient at random initial fields for systems of different $N$ with the same type of control Hamiltonian structure: a linearly increasing ${\cal G}_{max}$ with $N$ indicates minimal scaling with $N$, while sub-linear increase of ${\cal G}_{max}$ indicates exponential scaling with $N$. A more accurate prediction of the search effort can be made by measuring the ${\cal G}$ near, but not at, the optimum. Figure \ref{itergrad} shows the absolute search effort for searches using different $D$ structures plotted versus the measured value of \gr\ms at $J$=2 and $J$=0.01. Even at $J$=2, the gradient provides a good estimate of the absolute effort. The local curvature \curv\ms at the optimum is also an indicator of search effort scaling, as shown in Figure \ref{gc}(b). For control Hamiltonians that allow distant transitions and have sub-exponential scaling of effort with $N$ ($D$=1.0 and sparse structures), the curvature is flat as $N$ increases from 4 to 16. For control Hamiltonians that forbid distant transitions and have exponential scaling ($D<$1.0 and the tensor product structures), the curvature decreases with $N$ according to a power law (note the log-log plot). Under all circumstances, a smaller value of \curv\ms near the optimum indicates a greater search effort.

To understand the effects of Hessian local curvature on convergence efficiency, note that near a stable fixed point $\bar \e(t)$ of $J$ on the global optimum submanifold, (i.e., for $\norm{\e(t) - \bar \e(t)} \leq \epsilon$), the objective function is approximately quadratic in $\e(t)$, and we can linearize the differential equation (\ref{Egrad}) around $\bar \e(t)$ as
\begin{equation}\label{linearized}
\frac{\partial \e(s,t)}{\partial s} = -\int_0^T \frac{\delta^2 J}{\delta \e(s,t) \delta \e(s,t')}[\e(s,t') -\bar \e(t')]~dt'.
\end{equation}
To facilitate the convergence analysis, we assume that $J(\e) \rightarrow
J(\e)+\lambda \int_0^T L(\e(t))~dt$. In the limit $\lambda \rightarrow 0$, the cost functional is Mayer. For sufficiently small nonzero $\lambda$, i.e., $\lambda << 1$, the cost functional is Bolza and the optimal control problem has
a \emph{unique} solution $\bar \e(t)$. Then, the Hessian $\frac{\delta^2 J}{\delta \e(t) \delta \e(t')}$ is positive definite and by the Hartman-Grobman theorem for hyperbolic fixed points \cite{perko}, locally near the optimum, $\e(s,t)$ converges exponentially to $\bar \e(t)$ at a rate that is bounded by the smallest eigenvalue of the Hessian of the linearized system:
\begin{align*}
\norm{\e(s,t)-\bar \e(t)} &= \norm{\int_0^T \exp\left[-\frac{\delta^2 J}{\delta \e(s,t) \delta \e(s,t')}s\right](\e(0,t')-\bar \e(t'))~dt'}\\
&\leq \exp[-\lambda_{min} s]\norm{\e(0,t)-\bar \e(t)}.
\end{align*}
Since this eigenvalue $\lambda_{min}$ increases with local landscape curvature, higher curvature near the optimum facilitates convergence, for both the perturbed and original optimization problems.

\subsection{Complexity of Search Trajectories}\label{distance}
The ratio $R_{\e}$ measures the degree to which the search trajectory deviates from a direct path between the initial and final control fields. A statistical examination of $R_{\e}$ for the optimizations performed in Section \ref{scale1} shows that an increase in search effort with $N$ correlates with an increase in $R_{\e}$ with $N$ (not shown). Nevertheless, the ratio is always small ($R_{\e}<3$), indicating that while a linear trajectory from initial to final field cannot be followed, the trajectories followed are relatively direct.

\begin{figure}
\includegraphics[width=8.5cm]{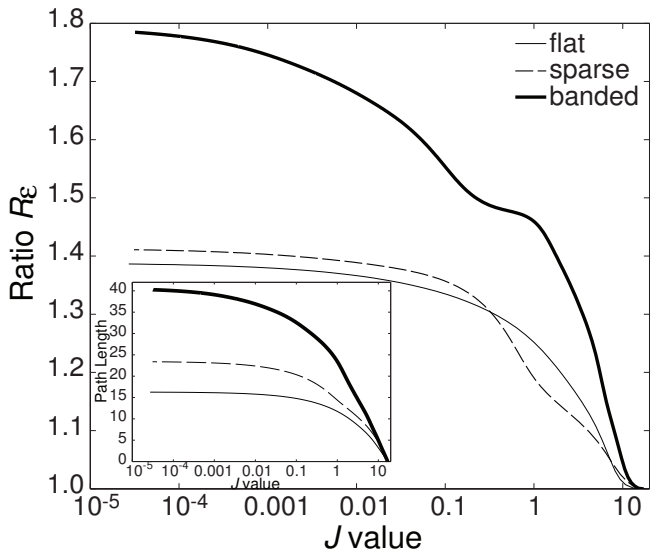}
\caption{Ratio $R_{\e}$ as a function of $J$ value for searches beginning from the same initial field with flat, sparse, and banded $\mu$ structures. The inset shows the path length for these searches as a function of $J$ value. \label{rat}}
\end{figure}

Further insight into the effect of the search trajectory on the required effort can be gained by examining the evolution of $R_{\e}$ over the search trajectory (i.e., with respect to the value of $J$). Three cases at $N$=8 starting from the same initial field (fluence $f$=1 with 10 evenly spaced frequency components) illustrate the difference in the complexity of the search trajectories. With the rotor $H_0$ structure, the control Hamiltonians used are (a) fully coupled (called ``flat'' here), (b) sparse with 50$\%$ allowed transitions, and (c) banded with two off-diagonal bands. The trajectories of the ratio $R_{\e}$ are plotted in Figure \ref{rat}. These results show that the optimization with the flat structure takes a direct path from initial to final field, while the optimizations with sparse and banded structures must change direction to optimize below $J$=1. In the vicinity of the optimum below $J$=0.1, the sparse $\mu$ optimization again can follow a direct path, while the trajectory for the banded $\mu$ continues to change direction.

\begin{figure}
\includegraphics[width=8.5cm]{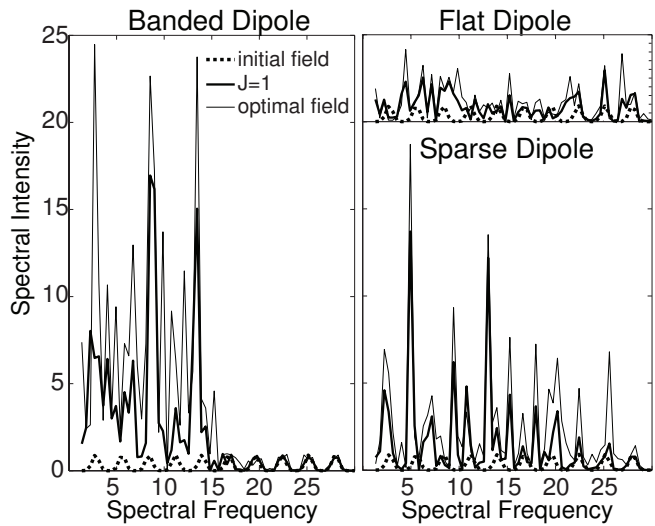}
\caption{Fourier spectra of initial field (dotted line), field at $J$=1 (thick line), and optimal field (thin line) for banded structure (left), sparse structure (bottom right), and flat structure (top right). The ordinate is of the same scale for all plots. All values are unitless.\label{ftcompare}}
\end{figure}

Examination of the Fourier spectra of the fields along the search trajectories reveals how field modes required for propagator control are progressively generated by local optimization. The spectra of the initial field, field at $J$=1, and optimal field for the three searches above are shown in Figure \ref{ftcompare}. At $J$=1, the fields have higher fluence and enhanced specific frequencies, particularly with low-frequency components for the banded structure since only near transitions are allowed. For the banded structure, many new frequencies are added or greatly enhanced when going from $J$=1 to the optimum. In contrast, all necessary frequencies are present at $J$=1 for searches with the sparse and flat $\mu$ structures.

\begin{figure}
\includegraphics[width=8.5cm]{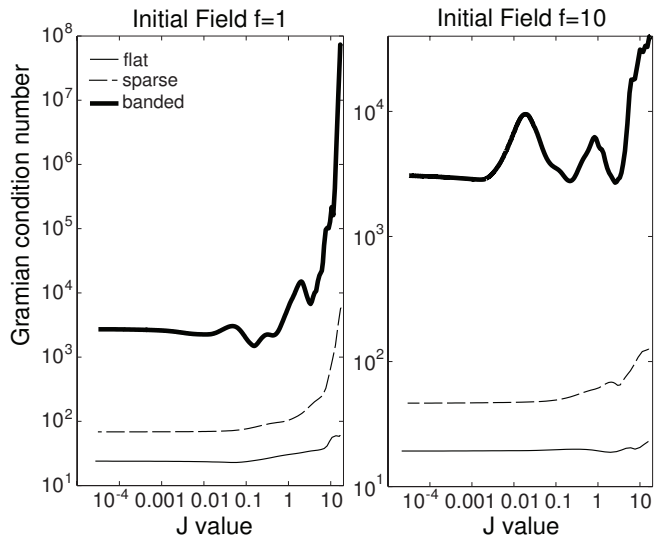}
\caption{Gramian matrix condition number versus $J$ value for searches beginning from the same initial field at $f$=1 (left) and $f$=10 (right) for the flat, banded, and sparse $\mu$ structures.\label{gmat}}
\end{figure}

The origin of the more complicated search trajectories and optimal fields for the banded $\mu$ structure can be explained by examining the condition number 
of the Gramian matrix (\ref{gramian}) along the search trajectory.
Consistent with the analysis in Section \ref{gramsection}, motion in certain directions on $\U(N)$ - such as those necessary to reach $W$ - is achieved more slowly for poorly conditioned Gramian matrices than for well-conditioned
Gramian matrices. The trajectories of the Gramian matrix condition number for the three searches above are shown in Figure \ref{gmat}(a). For comparison, the trajectories using an initial field of $f$=10 with the same frequencies is shown in Figure \ref{gmat}(b). The condition number remains orders of magnitude higher for the banded $\mu$ than for the sparse and flat $\mu$ throughout the search. Furthermore, the condition number levels off at a value under 100 for the sparse and flat $\mu$ structures, but remains well above 1000 for the banded structure, and displays more oscillations at $J<1$.

As discussed in Section \ref{gramsection}, the Gramian matrix is typically more well-conditioned for controls $\e(t)$ where all the terms in the Dyson series required for full controllability have become populated. Compared to the sparse $\mu$, the banded $\mu$ requires higher-order terms in the Dyson series to produce the desired $W$, and hence more frequency components in the optimal field. The flat $\mu$ structure requires the fewest Dyson terms, and thus shows the smallest difference between the fields at $J$=1 and the optimum. For a given distance $\norm{\e(s,t) - \bar \e(t)}$ from the global optimum, the accuracy of the linear approximation (\ref{linearized}) is greater for control systems with lower dynamical Lie algebra depth, due to lower order nonlinearity of the optimization problem. Away from critical points (outside the quadratic region), all terms in the Dyson series (\ref{dysonexp}) required to reach $W$ must be optimized, by the successive addition of new linear combinations of the real and imaginary
components of $\langle i | \mu(t) | j\rangle$ at each step.

\section{Conclusion}\label{conclusion}

We have provided a unifying picture of the convergence efficiency of first-order algorithms for unitary transformation control in terms of critical landscape topology and gradient flow dynamics. The roles of kinematic and system-dependent factors have been assessed. The results show that understanding the landscape topology is insufficient for predicting the required search effort to find an optimal field. Thus, in this work we have defined local landscape structure metrics based on a series expansion of the cost function variation with respect to the control, and have demonstrated that the first-order gradient-based metrics can qualitatively predict the required search effort. A central conclusion is that for control systems with low dynamical Lie algebra depth, the convergence efficiency is kinematically driven and any of the common first-order control optimization approaches based on unconstrained fields and a Mayer-type cost functional can be effective. In these cases, local gradient-based search algorithms can efficiently navigate the landscape for control of arbitrary unitary transformations, assuming that the system is controllable. In contrast, first-order algorithms are inefficient for systems of high Lie algebra depth. Future work should be aimed at quantitative classification of common gate control systems in terms of Lie algebra depth and identification of alternate search methods for systems of high Lie algebra depth.

The numerical results demonstrate that the control Hamiltonian structure determines the scaling of the required search effort with the Hilbert space dimension $N$ for optimization of arbitrary unitary transformations. In particular, control Hamiltonians $\mu$ that permit transitions between distant quantum states (e.g., the $D$=1 and sparse structures studied here) exhibit weak scaling of effort with $N$. For these systems, the first-order landscape structure metrics either grow linearly with $N$ (maximal gradient norm) or are invariant to $N$ (gradient norm near optimum), and second-order landscape structure metrics are invariant to $N$. The gradient flow was shown to be kinematically driven based on the Gramian matrix being well-conditioned throughout the search trajectory. Such systems have a low dynamical Lie algebra depth and are amenable to efficient first-order control optimization. In contrast, systems where transitions between distant states are weak or forbidden (e.g., $D<$1.0, banded, and tensor product structures) require an exponentially increasing search effort with $N$. The landscape structure metrics exhibit corresponding behavior, with the maximal gradient norm scaling sub-linearly with $N$ and the gradient norm near the optimum and second order metrics decreasing exponentially with $N$. Such systems have a greater dynamical Lie algebra depth, requiring higher-order terms in the Dyson expansion and exhibiting less well-conditioned Gramian matrices. Optimizations with these systems deviate from expected kinematic behavior, indicating that dynamical effects drive the gradient flow.

The results here suggest that for optimally controlling quantum gates, it is necessary to consider features of the control system other than controllability when engineering the time-independent Hamiltonian. In particular, the observation of exponential scaling with Hilbert space dimension for the $D$ and tensor product structures suggests that Lie algebra depth of $H_0$ and $\mu$ should be considered, with the engineering goal being to make transitions between distant quantum states allowed. Such Hamiltonian design efforts may be facilitated by the methods of Hamiltonian morphing \cite{vinny}, which allow the control Hamiltonian to be continuously deformed while holding the gate fidelity and the control field fixed.

Although the control of arbitrary unitary propagators is of fundamental importance, the primary focus of OCT studies of propagator control is for specific applications to quantum information sciences. Systems for which the landscape search complexity and resource scaling are favorable may be particularly useful for directly implementing multiqubit operations rather than decomposing them into sequences of one- and two-qubit universal gates. Search complexity for optimal control of multiqubit gates may thus be mitigated by choosing quantum information processing implementations where the control Hamiltonian can be tuned by design, with the goal being to produce control Hamiltonians that allow transitions between distant states. An example is quantum computation with polar molecule arrays in a magneto-optical trap \cite{demille}, where photoassociation techniques can be used to assemble novel atomic (e.g., homonuclear and heteronuclear alkali metal) dimers with differing permanent dipole moments. In such implementations the static electric field gradient that renders the molecules individually addressable can be used to orient the molecules so that the dipole-dipole coupling can be tuned, and qubits can be encoded on either ground or excited rovibrational states. Investigation of the scope of possible multiqubit control Hamiltonian structures accessible using such methods, and the application of OCT to these systems, is motivated by the present work.

\section*{Acknowledgments}
The authors acknowledge financial support from the Department of Energy, grant number DE-FG02-02ER15344. KWM acknowledges the support of a National Science Foundation graduate research fellowship.

\appendix

\section{Stable fixed point topology of PMP-iterative propagator control algorithms}\label{pmpiter}
The first-order algorithms below can only be formulated in discrete algorithmic time \cite{Ohtsuki}. A basic PMP-iterative algorithm proceeds via the following steps:
\begin{align*}
i\frac{d}{dt} \phi_k(t) &= (H_0 - \mu \cdot \tilde \e_k(t)) \phi_k(t),~\phi_k(T) = \nabla F(U_{k-1}(T))\\
i \frac{d}{dt} U_k(t) &= (H_0 - \mu \cdot \tilde \e_k(t)) U_k(t),~U_k(0)=U_0(0)
\end{align*}
where the costate equation is propagated backwards in time, with
\begin{align}
\tilde \e_k(t) &= \alpha_k \e_{k-1}(t) + \beta_k \langle \phi_k(t) | \mu | U_{k-1}(t)\rangle= \alpha_k \e_{k-1}(t) - \beta_k \tr\left[ i U_k^{\dag}(T) \nabla F(U_{k-1}(T)) U_k^{\dag}(t)\mu U_{k-1}(t) \right]\nonumber\\
\e_k(t) &= \alpha_k\tilde \e_k(t) + \beta_k\langle \phi_k(t) | \mu | U_k(t) \rangle \label{eupdate}
\end{align}

The constants, $\alpha_k \in [0,1]$ and $\beta_k < 0$ (for minimization of $J$), can in principle be chosen to be different in the $\tilde \e_k(t),\e_k(t)$ updates \cite{Ohtsuki}. 
The assignments of the constants $\alpha,\beta$ determine which type of cost functional $J$ is optimized by the algorithm.
In particular, the following values of $\alpha$ are of interest\footnote{Other choices for $\alpha_k,\beta_k$ - or modifications to equation (\ref{eupdate}) - can be used to optimize
either Bolza or Mayer costs, and are often required to ensure monotonic convergence of the algorithm, as discussed in \cite{Ohtsuki}. In particular, if $\alpha_k,\beta_k$ are selected outside of the intervals above, their values may
not be independent.}:
\begin{enumerate}
\item $\alpha_k=0,~\beta_k < 0$ minimizes \emph{Bolza} cost $J=F(U(T))+\frac{1}{2}\int_0^T \e^2(t)~dt$
\item $\alpha_k=1,~\beta_k < 0$ minimizes Mayer cost (\ref{cost})
\end{enumerate}
Iterative algorithms of type 2 have been applied \cite{tesch2001,Kosloff2002,Kosloff2003} to the problem of optimal gate control.
\footnote{Unlike homotopy algorithms, neither type of iterative algorithm introduced thus far can minimize field fluence while reaching high gate fidelity.}.

Prior work has demonstrated that PMP-iterative algorithms for quantum control converge monotonically
(i.e., $\delta J_{k,k-1} \leq 0$ at each step).  However, neither convergence to a global versus local optimum,
nor the rate of convergence were studied. The fixed points of type 2 discrete time PMP-iterative algorithms are points on the control landscape where $\e_k-\tilde \e_k=0$ or $\tilde \e_k-\e_{k-1}=0$ ($\langle \phi_k(t) | \mu | U_k(t) \rangle=0$ or $\langle \phi_k(t) | \mu | U_{k-1}(t) \rangle=0$).  In order for all such points to lie on the critical manifolds identified in Section 2.2, we must require that the Gramian matrices
\begin{align*}
\int_0^T \nu\left[U_k(T) U_k^{\dag}(t)\mu U_{k-1}(t)\right] \nu^T\left[U_{k-1}^{\dag}(t)\mu U_{k}(t)U^{\dag} _k(T)\right]~dt,~~\int_0^T \nu\left[U_k(T) \mu_k(t)\right] \nu^T\left[\mu_k(t)U^{\dag}_k(T)\right]~dt
\end{align*}
are nonsingular at successive steps of the algorithm. Then, the only fixed points of the discrete time
dynamical system correspond to points $U_k$ where $\nabla_U F(U(T))=0$.
Thus, if the HQF in Section \ref{kinematics} is positive definite at $U_k(T)=\hat U$,
then $\underset{k \rightarrow \infty}{\lim} U_k(T) = \hat U$, for some $\epsilon$ that is
equal to the radius of the attracting region of the critical point $\hat U$.
As shown in Section \ref{gradalgs}, the only critical point of $F(U)$ that satisfies
this criterion for asymptotic convergence is $\hat U = W$ and the associated neutrally stable controls $\bar \e(t)$
lie on the global minimum submanifold of $J(\e(\cdot))$. 

\section{Hessian quadratic form and rank}\label{hess}

The explicit form of the matrix $A^2$ in the HQF expression (\ref{hqf}) can be obtained from the second variation in the Taylor expansion of $J(\e +\delta\e)$; we find
\begin{equation*}
A^2=\int_0^T \int_0^T \delta \e(t)\cdot \mu(t)\mu(t')\cdot \delta \e(t') ~dt~dt'.
\end{equation*}
Assuming that the Gramian (\ref{gramian}) is nonsingular, i.e., that the real and imaginary components of the elements of $\mu(t)$ are linearly independent functions of time, $A$ can be any Hermitian matrix with associated direction $U(T)A$ in the tangent space $\mathcal{T}_U\U(N)$ to the unitary group \footnote{Note the third term in equation (\ref{hessiankernel}) does not contribute to the second order variation and hence does not have a corresponding term in the HQF; this is consistent with the fact that the second order variation is a quadratic form only at critical points, and cannot be used to assess the definiteness of the Hessian away from such points.}.

The range of $\mathcal{H}$ is spanned by eigenfunctions of the Hessian kernel:
these eigenfunctions $f_{\nu}(t)$, which satisfy $\int_0^T \mathcal{H}(t,t')f_{\nu}(t')dt' =  \lambda_{\nu} f_{\nu}(t)$,
are linear combinations of products of the real and imaginary components
$\Re\langle i | \mu(t)| j \rangle,\Im\langle i | \mu(t)| j \rangle$ of the time-evolved dipole operator, i.e., 
\begin{equation}\label{hessianmu}
\mathcal{H}(t,t') = \left(\sum_{j\geq i}a_{ij} \Re{\langle i | \mu(t)| j \rangle}+a_{ij}' \Im{\langle i | \mu(t)| j \rangle}\right)\left(\sum_{l\geq k}a_{kl}\Re{\langle k | \mu(t') | l \rangle}+a_{kl}'\Im{\langle k | \mu(t') | l \rangle}\right),
\end{equation}
where the expansion coefficients can be computed from equation (\ref{hessiankernel}).
This immediately implies that $\mathcal{H}(t,t')$ is a finite rank kernel, with
$\mathrm{rank} ~\mathcal{H}(t,t') \leq N^2$, even away from critical points where the HQF in Section 2
cannot be used to assess rank. 

\section{Numerical details}\label{num}

The control field $\e(t)$ was discretized on a time interval $t\in[0,T]$ in arbitrary dimensionless units into a sufficient number of time points to resolve the $|1\rangle\to|N\rangle$ transition frequency in $H_0$. For the rotor Hamiltonian (\ref{ho}) with $T$=14, 512 points were used for $N\leq 8$, 2048 points for $N$=16, and 4096 points for $N$=32. For the oscillator Hamiltonian (\ref{osc}), 512 points were used for $N\leq 8$, 1024 points for $N$=16, and 2048 points for $N$=32 were used. When $T>14$, 4096 points were used. For the simulations in Sections \ref{scale}, the initial control field $\e(0,t)$ contained $K$=20 Fourier components randomly chosen from a uniform distribution on an interval $[0, \omega_{1N}]$, where $\omega_{1N}$ denotes the $|1\rangle\to|N\rangle$ transition frequency.

\begin{figure}
\includegraphics[width=8.5cm]{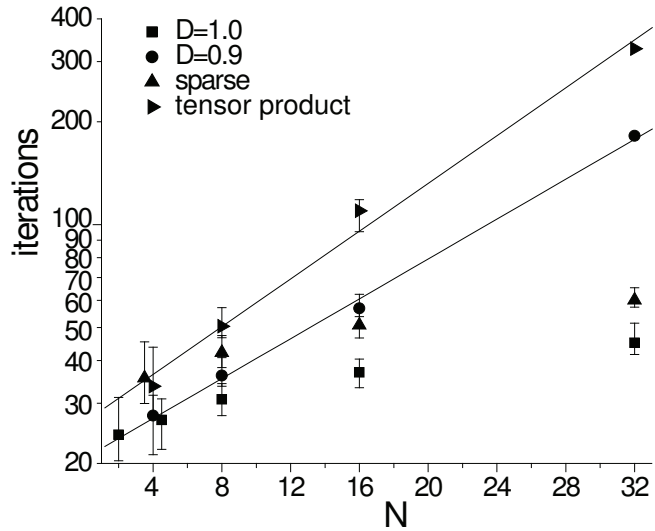}
\caption{Mean search effort (algorithmic iterations) with left and right standard deviation versus $N$ for rotor $H_0$ with $\mu$ structures $D$=1.0 (squares), $D$=0.9 (circles), sparse (up triangles), and tensor product (right triangles).\label{deffort}}
\end{figure}

\begin{figure}
 \includegraphics[width=8.5cm]{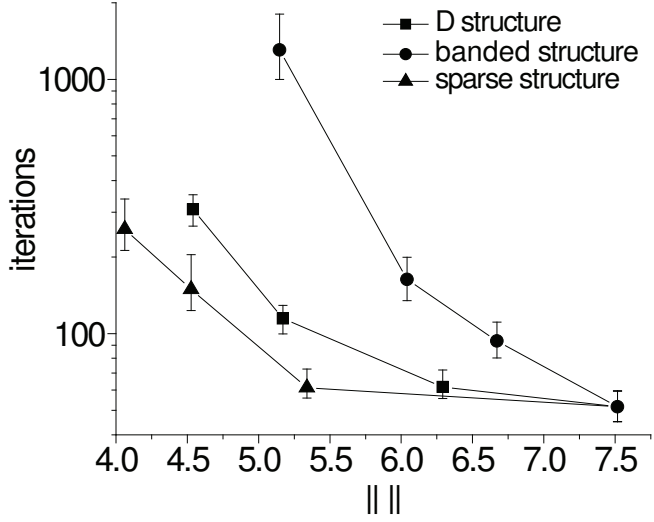}
\caption{Mean search effort versus norm $||\mu||$ for $N$=8 with control Hamiltonian structures $D$ (c.f. Eq. \ref{mu}, squares), sparse (triangles), and banded (circles).\label{dip}}
\end{figure}

Reported search effort is the number of RK4 algorithm iterations required to attain a $J$ value below the convergence criterion. This was $J\leq 10^{-6}\times J_{max}$ for all simulations except those in Section \ref{scale1}, where the criterion was $J\leq 10^{-3}\times J_{max}$. The sample size to generate the reported statistics was 1000 for the simulations in Section \ref{saddles}, and 20 for the simulations in Sections \ref{scale1}, \ref{scaled}, and \ref{searchstruct}. In Tables 1 and 2, the mean effort and standard deviation are reported. For the particularly difficult optimizations shown in Table 2 reporting no standard deviation, only one search was performed. In Figures \ref{deffort} and \ref{dip}, the error bars report the left and right standard deviations.

The metrics in Section \ref{str} were calculated by approximating the integrals as sums over the discretized time-points.

\renewcommand{\baselinestretch}{1}
\newpage
\begin{table}
 \begin{tabular}{|c|c|c|r|r|r|r|r|}
  \hline
{\bf N}&{\bf D}&{\bf W}&{\it f}&{\bf effort}&${\cal S}<0.1$ &${\cal S}<0.05$ &${\cal S}<0.01$\\\hline
2&1.0&$I_N$&10&31.3$\pm$16.1&0.193 (36.5)&0.102 (40.6)&0.030 (50.7)\\\hline
&&&0.1&32.6$\pm$2.2&0&&\\\hline
&&random&10&36.8$\pm$15.7&0.180 (38.5)&0.108 (39.5)&0.026 (42.5)\\\hline
&&&0.1&28.5$\pm$2.8&0.244 (29.9)&0.135 (30.4)&0.038 (31.8)\\\hline
&&FT&10&39.7$\pm$16.6&0.198 (40.6)&0.108 (42.4)&0.031 (47.3)\\\hline
&&&0.1&23.9$\pm$3.4&0.110 (27.7)&0.028 (29.0)&0\\\hline
4&1.0&$I_N$&10&43.9$\pm$15.6&0.081 (52.6)&0.027 (53.4)&0.003 (43)\\\hline
&&&0.1&47.2$\pm$8.2&0.076 (63.6)&0.036 (78.6)&0.002 (86)\\\hline
&&random&10&40.6$\pm$10.1&0.078 (46.4)&0.028 (50.5)&0.002 (52)\\\hline
&&&0.1&38.7$\pm$6.9&0.032 (43.4)&0.013 (46.7)&0\\\hline
&&FT&10&47.5$\pm$19.3&0.085 (52.6)&0.039 (53)&0.003 (59)\\\hline
&&&0.1&45.9$\pm$13.8&0.277 (46.5)&0.059 (48.2)&0\\\hline
&0.6&FT&10&51.1$\pm$13.5&0.043 (58.5)&0.012 (58.6)&0.001 (61)\\\hline
&&&0.1&45.1$\pm$8.9&0.044 (50.1)&0.016 (53.4)&0.003 (62)\\\hline
&&random&10&41.9$\pm$11.8&0.043 (58.5)&0.012 (58.6)&0.001 (61)\\\hline
&&&0.1&37.7$\pm$7.0&0.033 (43)&0.019 (43.5)&0.001 (35)\\\hline
8&1.0&$I_N$&10&49.9$\pm$6.7&0.027 (55.2)&0.009 (57.4)&0\\\hline
&&&0.1&74.9$\pm$13.8&0.010 (84.4)&0.005 (83.2)&0\\\hline
&&random&10&46.8$\pm$5.6&0.022 (54.2)&0.006 (53)&0\\\hline
&&&0.1&57.9$\pm$5.3&0.005 (64.2)&0.001 (61)&0\\\hline
&&FT&10&48.1$\pm$5.9&0.027 (53.7)&0.008 (53.1)&0\\\hline
&&&0.1&59.9$\pm$6.4&0.109 (63.3)&0.016 (66.3)&0\\\hline
 \end{tabular}
\caption{Search effort mean/standard deviation and probability of encountering saddles. For the column labeled $W$, $I_N$ denotes the $N$-dimensional identity matrix and FT denotes the quantum Fourier transform gate. $f$ denotes the initial field strength. Effort is the number of iterations required to achieve $J<10^{-6}\times J_{max}$. The last three columns show the fraction of searches to encounter saddle manifolds to within $\cal S$ values of 0.1, 0.05, and 0.01; the numbers in parentheses denote the mean search effort for searches that encountered these saddles.}
\label{saddletable}
\end{table}

\begin{table}[t]
 \begin{tabular}{|r|c|c|c|c|c|c|c|}
\hline
{\it N}&$\mu$&{$H_0$}&Effort, $f$=10&Effort, $f$=$10^{-4}$&Effort, $f$=10$^3$&{$ H_0$}&Effort, $f$=10\\\hline\hline
2&D=1.0&rotor&24.3$\pm$6.3&15.4$\pm$2.5&106$\pm$65&oscillator&16.4$\pm$2.5\\\hline
4&D=1.0&rotor&26.8$\pm$4.5&31.4$\pm$5.8&182$\pm$120&oscillator&24.1$\pm$2.8\\\hline
8&D=1.0&rotor&30.8$\pm$3.3&38.4$\pm$4.4&54.4$\pm$9.5&oscillator&23.9$\pm$2.5\\\hline
16&D=1.0&rotor&36.9$\pm$3.5&45.1$\pm$2.6&35.8$\pm$3.1&oscillator&26.7$\pm$2.7\\\hline
32&D=1.0&rotor&45.1$\pm$4.9&&&oscillator&53.9$\pm$3.3\\\hline
4&D=0.9&rotor&27.6$\pm$5.2&28.1$\pm$4.6&80.1$\pm$48.1&oscillator&25.6$\pm$4.8\\\hline
8&D=0.9&rotor&36.2$\pm$3.6&41.1$\pm$5.5&56.8$\pm$8.9&oscillator&29.7$\pm$2.7\\\hline
16&D=0.9&rotor&56.9$\pm$4.4&64.2$\pm$4.7&58.7$\pm$6.8&oscillator&40.3$\pm$4.1\\\hline
32&D=0.9&rotor&182&&&oscillator&358\\\hline
4&D=0.6&rotor&37.0$\pm$6.9&32.8$\pm$6.5&66.9$\pm$18.1&oscillator&34.5$\pm$6.9\\\hline
8&D=0.6&rotor&90.4$\pm$17.3&94.2$\pm$14.3&116$\pm$29&oscillator&90.9$\pm$7.7\\\hline
16&D=0.6&rotor&2169&&1040&oscillator&651$\pm$134\\\hline
4&sparse&rotor&35.6$\pm$7.7&140$\pm$73&120$\pm$52&oscillator&31.3$\pm$5.9\\\hline
8&sparse&rotor&42.3$\pm$4.6&58.6$\pm$6.9&60.8$\pm$9.3&oscillator&42.9$\pm$4.6\\\hline
16&sparse&rotor&50.8$\pm$4.6&70.1$\pm$4.8&45.9$\pm$3.2&oscillator&41.9$\pm$2.8\\\hline
32&sparse&rotor&60.2$\pm$3.9&&&oscillator&77.2$\pm$1.9\\\hline\hline
N&$\mu$&$H_0$&effort, $f$=10&$H_0$&effort, $f$=10&&\\\hline
4&tensor product&rotor&33.7$\pm$6.2&oscillator&27.5$\pm$4.8&&\\\hline
8&tensor product&rotor&50.4$\pm$5.3&oscillator&61.2$\pm$9.4&&\\\hline
16&tensor product&rotor&109.8$\pm$10.8&oscillator&143.6$\pm$15.5&&\\\hline
32&tensor product&rotor&327&oscillator&916&&\\\hline\hline
{\it N}&$\mu$&{$H_0$}&Effort, $f$=10&comment&&&\\\hline
2&D=1.0&rotor&27.5$\pm$7.7&FT gate&&&\\\hline
4&sparse&rotor&30.8$\pm$7.9&FT gate&&&\\\hline
8&sparse&rotor&49.7$\pm$4.7&FT gate&&&\\\hline
16&sparse&rotor&50.5$\pm$4.7&FT gate&&&\\\hline
32&sparse&rotor&61.9$\pm$3.6&FT gate&&&\\\hline
4&D=0.6&rotor&20.8$\pm$2.7&T=28&&&\\\hline
8&D=0.6&rotor&39.0$\pm$6.7&T=56&&&\\\hline
16&D=0.6&rotor&234$\pm$30&T=112&&&\\\hline
\end{tabular}
\caption{Scaling of search effort with $N$ for different choices of $H_0$, $\mu$ structure, initial field strength $f$, time $T$, and target $W$.}
\label{scaletable}
\end{table}

\newpage

\end{document}